\documentclass[12pt]{article}

\usepackage{scicite}
\usepackage{times}
\usepackage{amsmath,amssymb,graphicx,fullpage,color,mathtools,amsthm,xcolor}
\usepackage{caption,subcaption}
\usepackage{algorithm,algorithmic}
\usepackage{cite}
\usepackage{microtype}
\usepackage{hyperref}
\usepackage{amsbsy}
\usepackage{enumerate}
\usepackage{url}
\usepackage{amsthm}
\usepackage[CaptionAfterwards]{fltpage}
\theoremstyle{plain}
\newtheorem*{theorem*}{Theorem}

\definecolor{webgreen}{rgb}{0,.35,0}
\definecolor{webbrown}{rgb}{.6,0,0}
\definecolor{RoyalBlue}{rgb}{0,0,0.9}
\definecolor{purp}{rgb}{0.6,0.05,0.8}
\definecolor{ora}{rgb}{0.7,0.35,0.02}

\hypersetup{
   colorlinks=true, linktocpage=true, 
   urlcolor=webbrown, linkcolor=RoyalBlue, citecolor=webgreen,
   pdfauthor={Gary P. T. Choi, Levi H. Dudte, L. Mahadevan},
   pdfsubject={Compact reconfigurable kirigami}
}

\newenvironment{sciabstract}{%
\begin{quote} \bf}
{\end{quote}}

\newcounter{lastnote}

\begin{document}
\author{Gary P. T. Choi$^{1}$, Levi H. Dudte$^{2}$, L. Mahadevan$^{2,3\ast}$\\
\\
\footnotesize{$^{1}$Department of Mathematics, Massachusetts Institute of Technology, Cambridge, MA, USA}\\
\footnotesize{$^{2}$School of Engineering and Applied Sciences, Harvard University, Cambridge, MA, USA}\\
\footnotesize{$^{3}$Departments of Physics, and Organismic and Evolutionary Biology, Harvard University, Cambridge, MA, USA}\\
\footnotesize{$^\ast$To whom correspondence should be addressed; E-mail: lmahadev@g.harvard.edu}
}
\title{Compact reconfigurable kirigami}
\date{} 

\baselineskip24pt

\maketitle

\begin{sciabstract}
Kirigami involves cutting a flat, thin sheet that allows it to morph from a closed, compact configuration into an open deployed structure via coordinated rotations of the internal tiles. By recognizing and generalizing the geometric constraints that enable this art form, we propose a design framework for compact reconfigurable kirigami patterns, which can morph from a closed and compact configuration into a deployed state conforming to any prescribed target shape, and subsequently be contracted into a different closed and compact configuration. We further establish a condition for producing kirigami patterns which are reconfigurable and rigid deployable allowing us to connect the compact states via a zero-energy family of deployed states. All together, our inverse design framework lays out a new path for the creation of shape-morphing material structures.
\end{sciabstract} 

Kirigami is the art of using cuts in a single sheet of paper that allow it to deform and change shape via coordinated rotations. In recent years, the idea of kirigami has motivated the design of metamaterials, where architected cuts on a flat, thin sheet of material can lead to special properties not found in most naturally occurring materials, such as a negative Poisson's ratio~\cite{grima2004negative}. There has been a vast number of studies on the geometry, topology and mechanics of kirigami
~\cite{mitschke2013finite,shan2015design,rafsanjani2016bistable,chen2016topological,choi2019programming,chen2020deterministic,jiang2020freeform}
%~\cite{mitschke2013finite,shan2015design,rafsanjani2016bistable,chen2016topological,tang2017design,tang2017programmable,rafsanjani2017buckling,moshe2019kirigami,rafsanjani2019propagation,lubbers2019excess,choi2019programming,chen2020deterministic,jiang2020freeform}, 
with applications to the design of nanocomposites~\cite{blees2015graphene,shyu2015kirigami}, shape-morphing sheets~\cite{neville2016shape,celli2018shape}, inflatable structures~\cite{konakovic2018rapid}, soft robots~\cite{rafsanjani2018kirigami} etc. 
%Closely related to the art of paper cutting is the art of paper folding, which is known as origami. In recent years, several studies have focused on the design of reconfigurable origami-inspired metamaterials that admit multiple folded states~\cite{silverberg2014using,sussman2015algorithmic,filipov2016origami,overvelde2017rational,fang2018reconfigurable,dieleman2020jigsaw,wang2020active}. 
Almost without exception, the deployed kirigami structures are open, and the property of admitting multiple closed and compact contracted states has been addressed only in a few well-known periodic regular kirigami patterns~\cite{lipton2018handedness,yang2018geometry,stavric2019geometrical}, and lead to states that are related to each other via global rotations. This raises a natural question: is it possible to introduce cuts on a thin sheet of material in a way that yields a deployable and reconfigurable kirigami structure conforming to any prescribed shape and admitting multiple closed and compact contracted states? Here we answer this question in the affirmative by building on our recent inverse design framework~\cite{choi2019programming} by identifying a set of geometric constraints to achieve both reconfigurability and a range of different energy landscapes associated with the deployment pathway.
%Closely related to the art of paper cutting is the art of paper folding, which is known as origami. In recent years, several studies have focused on the design of reconfigurable origami-inspired metamaterials, including reconfigurable Miura-ori tessellations~\cite{silverberg2014using}, honeycomb lattice with holes that can be folded into different shapes by varying the fold directions~\cite{sussman2015algorithmic}, origami tubes with reconfigurable polygonal cross sections~\cite{filipov2016origami}, regular prismatic structures with reconfigurable spatial architecture~\cite{overvelde2017rational}, and pluripotent crease patterns that are capable of folding into multiple shapes~\cite{dieleman2020jigsaw}. However, to the best of our knowledge, there has not been any study on the reconfigurability of kirigami tilings. It is natural to ask if it is possible to introduce cuts on a sheet of material in a way that yields a deployable structure admitting more than one closed and compact state. 

%Recently, we developed a novel inverse design framework for creating generalized kirigami patterns that can be deployed and conform approximately to any prescribed target shape in two or three dimensions~\cite{choi2019programming}. Building on this framework, here we pose and solve the problem of designing reconfigurable kirigami patterns and explore their design space. 

\begin{figure}[t!]
\centering
\includegraphics[width=\textwidth]{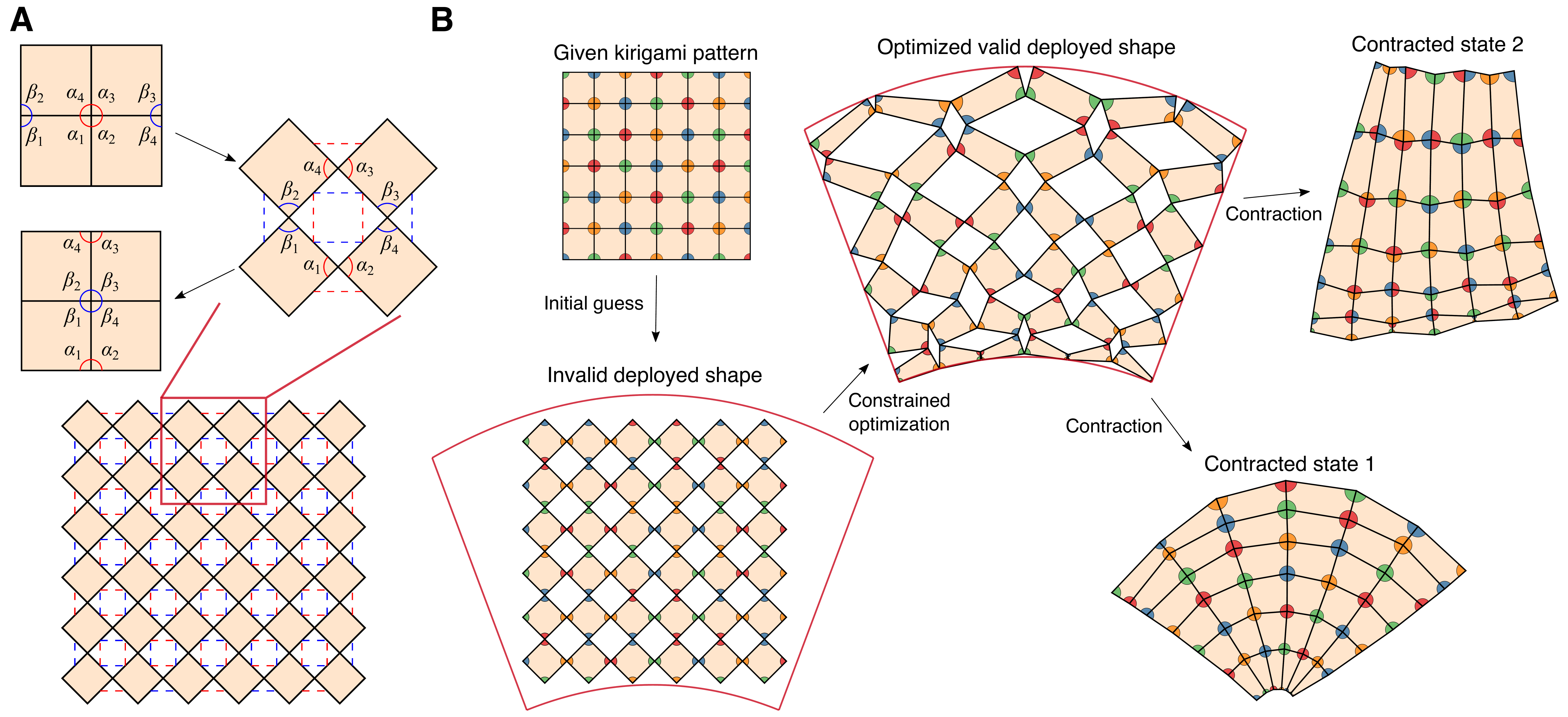}
\caption{\textbf{Reconfigurable kirigami design.} (A) An enlargement of the unit cell of a quad kirigami tessellation illustrating the constraints in edge lengths and angles to be satisfied (Eq.~\eqref{eqt:edge_length_constraint}--\eqref{eqt:dual_angle_sum_constraint}). The red dotted lines indicate the ordinary edge pairs corresponding the same cuts, and the blue dotted lines indicate the dual edge pairs for getting the other contracted configuration. (B) The inverse design framework for reconfigurable kirigami. Starting with a given kirigami pattern and a prescribed target shape, we construct an initial guess in the deployed space and solve a constrained optimization problem to obtain a valid deployed configuration that satisfies both the ordinary contractibility constraints and the new reconfigurability constraints and matches the target shape. We then contract the deployed configuration in two ways, one by following the cut edge pairs and one by following the dual edge pairs, and obtain two contracted states of it. The angles are colored based on the correspondence in the given kirigami pattern.}
\label{fig:F1}
\end{figure}

To crystallize our question in a minimal setting, we consider the quad kirigami patterns, a class of deployable structures obtained by introducing cuts to form quadrilateral tiles connected at hinges (see Fig.~\ref{fig:F1}A) with a single global degree of freedom. To determine the size and orientation of the cuts that yield a deployed configuration approximating a prescribed target shape, it is more convenient to work in the deployed space and change the geometry of each tile. In our recent work~\cite{choi2019programming}, we have shown that the key criteria for guaranteeing that the deformed deployed configuration yields a valid kirigami pattern, also known as the \textit{contractibility constraints}, consist of the \textit{edge length constraints} and the \textit{angle sum constraints}. As illustrated by the red dotted lines in Fig.~\ref{fig:F1}A, the edge length constraints enforce the constancy of the length of each pair of edges $\{\mathbf{e}_{i_1,1}, \mathbf{e}_{i_1,2}\}$ in the deployed space {that correspond to the same edge $\mathbf{e}_{i_1}$ in the pattern space}, so that
\begin{equation} \label{eqt:edge_length_constraint}
l_{i_1,1} = l_{i_1,2},
\end{equation}
where $l_{i_1,k} = \|\mathbf{e}_{i_1,k}\|$ is the length of the edge $\mathbf{e}_{i_1,k}$. The angle sum constraints enforce the condition that the set of angles $\{\alpha_{k}\}_{k=1}^n$ in the deployed space that correspond to the same vertex in the pattern space (see Fig.~\ref{fig:F1}A) satisfy
\begin{equation}\label{eqt:angle_sum_constraint}
\sum_{k=1}^n \alpha_{k} = 2\pi.
\end{equation}

To ensure that the deformed deployed configuration admits another closed and compact contracted state, we exploit an underlying duality present in the standard quad kirigami pattern associated with \textit{reconfigurability, which implies the presence of dual pairs of length and angle constraints.} These read as follows:
\begin{enumerate}[(i)]
\item (\textit{Dual edge length constraints}) For each pair of adjacent edges $\{\mathbf{e}_{i_2,1}, \mathbf{e}_{i_2,2}\}$ where $\mathbf{e}_{i_2,1}, \mathbf{e}_{i_2,2}$ belong to two different tiles and are not paired up in Eq.~\eqref{eqt:edge_length_constraint} (as illustrated by the blue dotted lines in Fig.~\ref{fig:F1}A), we should have
\begin{equation} \label{eqt:dual_edge_length_constraint}
l_{i_2,1} = l_{i_2,2},
\end{equation}
{noting that the edges $\mathbf{e}_{i_2,1}, \mathbf{e}_{i_2,2}$ will then correspond to the same edge $\mathbf{e}_{i_2}$ in the reconfigured pattern space.}
\item (\textit{Dual angle sum constraints}) For every set of angles $\{\beta_{k}\}_{k=1}^n$ which are dual to the set of angles $\{\alpha_{k}\}_{k=1}^n$ mentioned in Eq.~\eqref{eqt:angle_sum_constraint} inside a unit cell (see Fig.~\ref{fig:F1}A), we should have
\begin{equation} \label{eqt:dual_angle_sum_constraint}
\sum_{k=1}^n \beta_{k} = 2\pi,
\end{equation}
{noting that the angles $\beta_{1},\beta_{2},\dots,\beta_{n}$ will then correspond to the same vertex in the reconfigured pattern space.}
\end{enumerate} 

Altogether, for quad kirigami, the contractibility constraints and the reconfigurability constraints enforce that all edges around each hole in the deployed configuration must be equal in length, yielding a rhombus, and that all angles of the deformed tiles at two opposite corners of each rhombus hole should add up to $2\pi$ (see SI Section~S1).

In addition to the internal constraints, to further ensure that the deployed configuration conforms to a prescribed boundary shape, we must also enforce the following \textit{shape matching constraints}~\cite{choi2019programming} for every boundary node $\mathbf{p}_i$:
\begin{equation} \label{eqt:shape_matching}
\|\mathbf{p}_i - \mathbf{\widetilde{p}}_i \|^2 = 0,
\end{equation}
where $\mathbf{\widetilde{p}}_i$ is the projection of $\mathbf{p}_i$ onto the target boundary. 

Given the above constraints, we are now in a position to frame the inverse design framework for reconfigurable kirigami, shown in Fig.~\ref{fig:F1}B. Given a regular kirigami tessellation and a prescribed target shape, we start with an initial guess of the deployed configuration, which can either be a trivial deployment of the standard tessellation, a deformed configuration produced by a conformal/quasi-conformal map~\cite{choi2018linear,meng2016tempo}, or any other methods that preserve the number and connectedness of the tiles. Almost without exception, any initial guess will violate at least some of the constraints in contractibility, reconfigurability, or target shape matching. To obtain a valid reconfigurable kirigami pattern, we formulate a constrained optimization problem for the deployed configuration with all constraints above together with the \textit{non-overlap constraints} which prevent adjacent tiles from overlapping (see SI Section~S2). Here we simply adopt the objective function used in our previous work~\cite{choi2019programming} to yield a smooth shape change over the entire pattern (see SI Section~S2), noting that other choices are possible. Expressing all constraints and the objective function in terms of the coordinates of the nodes in the deployed configuration, we solve this optimization problem using the \texttt{fmincon} routine in MATLAB (see SI Section~S2). This yields a deformed deployed configuration that satisfies all constraints, from which we can obtain the two contracted states by rotating the tiles according to the two sets of edge correspondences.

\begin{figure}[t!]
\centering
\includegraphics[width=\textwidth]{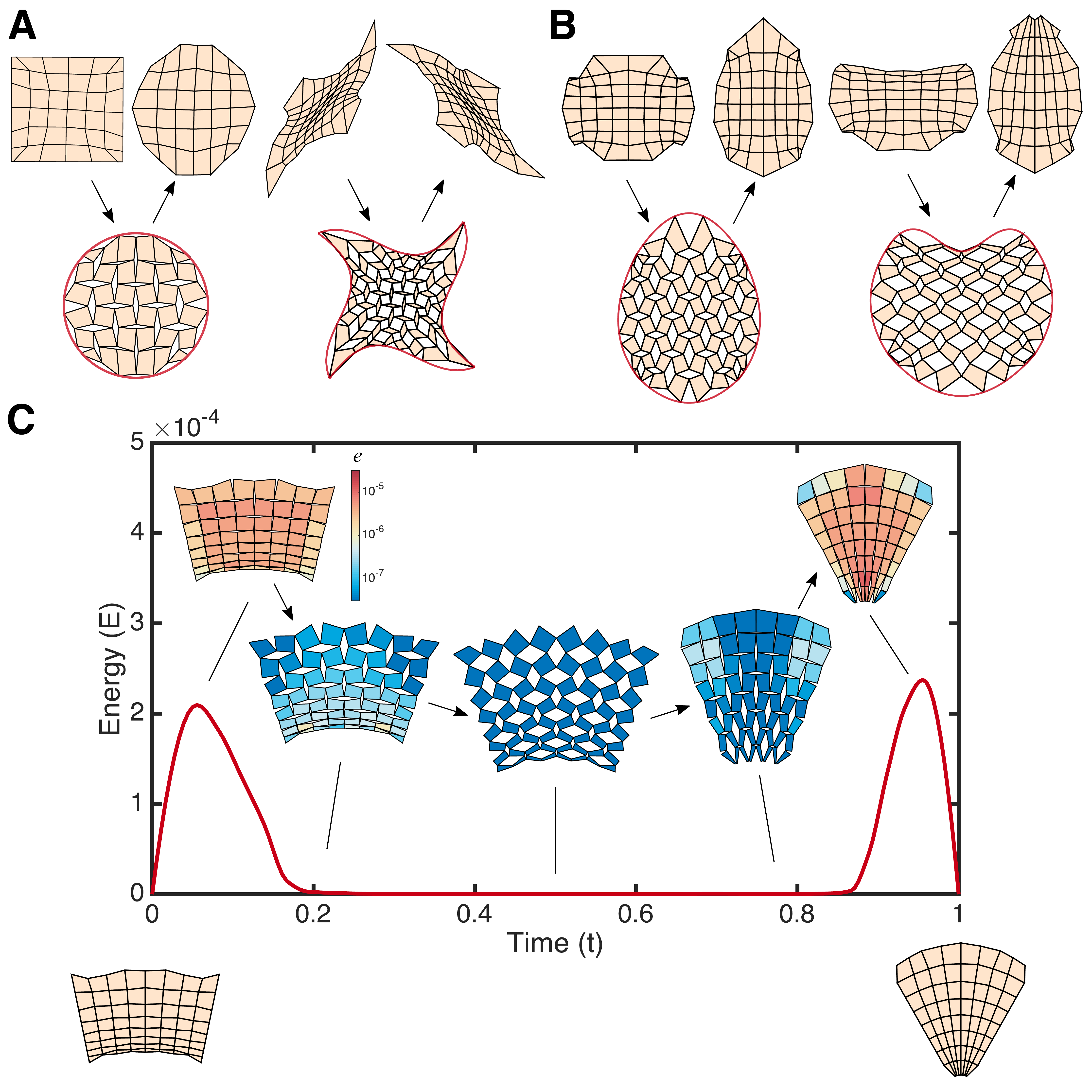}
\caption{\textbf{Reconfigurable kirigami patterns.} (A) Examples of reconfigurable kirigami patterns that conform to prescribed target shapes. For each example, the top row shows the two contracted states and the bottom row shows the deployed state. (B) Reconfigurable kirigami patterns produced by further enforcing a reflection symmetry constraint. (C) Energetics of the deployment and contraction of a reconfigurable kirigami pattern shows barriers near the two contracted states, but almost zero energy in between. This results in an unusual landscape - with monostable elastic minima at the ends and mechanism-like zero-energy in between. The insets show the intermediate deployed states with each tile color-coded by the total spring energy along all its edges and diagonals (denoted by $e$) (see video S1 and SI for details).}
\label{fig:F2}
\end{figure}

Fig.~\ref{fig:F2}A shows several examples of reconfigurable kirigami patterns obtained by our method, where each of the kirigami patterns admits two distinct contracted states and the deployed configuration conforms to a prescribed intermediate target shape. We see that our method is capable of approximating target shapes with different curvature properties, like our previous inverse design framework~\cite{choi2019programming}. Additionally, we can also control the boundary shape of a contracted state by introducing additional constraints on the boundary edge lengths and angles, yielding a reconfigurable kirigami pattern that deploys from a contracted rectangle to a circle and then contracts to another shape. {Of particular interest is the fact that it is possible to use microscopic tile rotations that induce local topological rearrangements to induce an effective overall global rotation.} In addition, if the deployed target shape is symmetric, one can further enforce this as an additional constraint in the optimization framework to produce reconfigurable kirigami patterns that are symmetric in the contracted and deployed states (Fig.~\ref{fig:F2}B). 

The presence of multiple closed and compact contracted states in these reconfigurable structures naturally implies the presence of at least a bistable mechanical energy landscape (Fig.~\ref{fig:F2}C), with additional minima arising as a function of new constraints. To characterize this energy landscape, we quantify the deformation of the tiles by replacing the edges and diagonals of the quads by linear springs with a total energy that reads (see SI Section~S4 for more details):
\begin{equation}\label{eqt:deployment}
    E(t) = \sum_{i,j: [i,j] \text{ is an edge or a diagonal}} \left(\frac{\|\mathbf{x}_i(t) - \mathbf{x}_j(t)\| - l_{ij}}{l_{ij}}\right)^2,
\end{equation}
where $\mathbf{x}_i(t)$ is the position of the node $i$ at time $t$, and $l_{ij}$ is the rest length of the spring at $[i,j]$. When we calculate the elastic energy as a function of the deployment stage, we find that the two compact end states are indeed global energy minima, but the tiles have to deformed to move away from these states. Interestingly, there is a regime of deployment where the system is more like a mechanism with the tiles essentially responding by just rotating (see Fig.~\ref{fig:F2}C inset and SI Video~S1). This is a generic feature (see SI Section~S5 for more results) of reconfigurable kirigami patterns that behave like elastic structures near their compact states, and like rigid mechanisms away from them.

Our previous example shows that generic quad kirigami patterns are not single degree-of-freedom mechanisms~\cite{choi2019programming}, and thus neither are reconfigurable kirigami patterns that we have introduced so far. This raises the natural question: how can we complement the geometric constraints in Eq.~\eqref{eqt:dual_edge_length_constraint} and Eq.~\eqref{eqt:dual_angle_sum_constraint} to preserve the single degree-of-freedom in the kirigami patterns? Said differently, how can make quad kirigami rigid-deployable by admitting a single continuous path from one contracted pattern through the solved deployed state to the second contracted pattern state such that all the constituent tiles rotate rigidly without deforming? To enable this, we introduce the following \textit{rigid-deployability constraints}: around every negative space (i.e. a hole formed by four edges of four neighboring quads), we should have
\begin{equation}\label{eqt:straight}
%    \theta_{i_1,1} + \theta_{i_1,2} = \theta_{i_1,3} + \theta_{i_1,4}  = \theta_{i_2,1}+ \theta_{i_2,2} = \theta_{i_2,3}+ \theta_{i_2,4} = \pi,
    \alpha_{1} + \alpha_{2} = \alpha_{3} + \alpha_{4}  = \beta_{1}+ \beta_{2} = \beta_{3}+ \beta_{4} = \pi,
\end{equation}
where the design angles are as shown in Fig.~\ref{fig:F1}. A mathematical justification of this set of constraints is provided below.

{\bf Lemma.} \textit{(Local rigid-deployability) A reconfigurable kirigami pattern is locally rigid-deployable if and only if the constraints in Eq.~\eqref{eqt:straight} are satisfied for all negative spaces.}

{\bf Proof.} Eq.~\eqref{eqt:straight} ensures that each negative space forms a straight line in both contracted configurations. Taken in isolation, each negative space can be thought of as a four-bar linkage (highlighted in blue in Fig.~\ref{fig:F3}A, left). A negative space/four-bar linkage from a generic quad kirigami pattern (i.e. one that is not reconfigurable) has two unique edge lengths where edges with equal lengths are incident to each other (see SI Fig.~S1). Such a four-bar linkage has two one-dimensional deployment paths in the plane connected to each other at two branch points, where the edges with equal lengths coincide with each other and all edges are collinear. In the plane, the four-bar linkage cannot move from one deployment path to another except at and through a branch point. Thus, quad kirigami patterns which do not satisfy the rigid-deployability constraints contain negative spaces which cannot pass from pattern to deployed states in the plane without changing edge lengths. And, conversely, quad kirigami patterns which satisfy the rigid-deployability constraints have only negative spaces which can rigidly deform from their straight-line pattern configurations to their solved, deployed configurations in the plane. Reconfigurable quad kirigami structures have negative spaces/four-bar linkages with all lengths being equal (see SI Fig.~S2). Such linkages have three one-dimensional deployment paths, one path in which all hinges are activated and the linkage forms a rhombus and two degenerate paths in which two of the four hinges in the linkage are activated, each connected to the rhombus path at a respective branch point (Fig.~\ref{fig:F3}A, right). Thus, reconfigurable quad kirigami patterns satisfying Eq.~\eqref{eqt:straight} have only negative spaces which can rigidly deform from their two straight-line pattern configurations to their solved, deployed configurations in the plane and hence are locally rigid-deployable. And thus if Eq.~\eqref{eqt:straight} is violated for some negative space in a reconfigurable quad kirigami pattern, the pattern cannot be locally rigid-deployable as the four-bar linkage cannot rigidly move to a branch point while remaining embedded in two dimensions.\hfill $\blacksquare$

After establishing the above lemma, we can prove the following result:
\begin{theorem*}
(Global rigid-deployability) A reconfigurable kirigami pattern is globally rigid-deployable if and only if the constraints in Eq.~\eqref{eqt:straight} are satisfied for all negative spaces.
\end{theorem*}
{\bf Proof.} The above lemma provides local rigid-deployability if and only if the constraints in Eq.~\eqref{eqt:straight} are satisfied for each negative space in the pattern. To analyze global rigid-deployability, we construct a loop condition $F$ around a single interior face in a generic (i.e. not necessarily reconfigurable) quad kirigami which must be identity at all points along a rigid-deployment. As shown in Fig.~\ref{fig:F3}A, let $\theta_{i,j}$ be design angles and $\phi_{i,j}$ be deployment angles in a quad kirigami four-bar linkage negative space. Let $f_i$ be the function that transfers a deployment angle $\phi_{i,1}$ to the deployment angle $\phi_{i+1,1}$ by composing angle-sum transfer $h_i$ and four-bar kinematics transfer $f_i$ such that $
    \phi_{i+1,1} = f_i(\phi_{i,1}) = g_i(h_i(\phi_{i,1})),
    \phi_{i,2} = h_i(\phi_{i,1}) = 2\pi - \phi_{i,1} - \theta_{i,1} - \theta_{i,2},
    \phi_{i+1,1} = g_i(\phi_{i,2}) = 2\sin^{-1}\left(\frac{l_{i,2}\sin\phi_{i,2}}{\sqrt{l_{i,1}^2 + l_{i,2}^2 - 2l_{i,1}l_{i,2}\cos\phi_{i,2}}}\right).
$
If the loop condition 
\begin{equation}
    F(\phi_{1,1}) = f_4(f_3(f_2(f_1((\phi_{1,1}))))) = \phi_{1,1}
\end{equation}
is satisfied for every value of $\phi_{1,1} \in [0,2\pi - \theta_{1,1} - \theta_{1,2}]$ for every interior quad, then the quad kirigami pattern is globally rigid-deployable. In a reconfigurable quad kirigami pattern, we have $\theta_{i,1} + \theta_{i,2} = \pi$ and $l_{i,1} = l_{i,2}$ and hence $
    \phi_{i,2} = h_i(\phi_{i,1}) = \pi - \phi_{i,1},
    \phi_{i+1,1} = g_i(\phi_{i,2}) = \pi - \phi_{i,2},
    \phi_{i+1,1} = f_i(\phi_{i,1}) = \phi_{i,1}.
$
So $F$ is a composition of identity functions $f_i$ and is itself identity. Thus, reconfigurable quad kirigami patterns satisfying Eq.~\eqref{eqt:straight} are globally rigid-deployable. \hfill $\blacksquare$

Therefore, we can obtain reconfigurable kirigami patterns which are (globally) rigid-deployable by simply augmenting our constrained optimization framework with the additional condition~\eqref{eqt:straight}.

%Fig.~\ref{fig:F4}A shows examples of reconfigurable and rigid-deployable kirigami patterns obtained by our method, from which it can be observed that our method is capable of producing a wide range of patterns to approximate different shapes even after enforcing the additional rigid-deployability constraints. 

Fig.~\ref{fig:F3}B shows a reconfigurable, rigid-deployable kirigami pattern obtained by our method. In contrast to the patterns in Fig.~\ref{fig:F2}, here each four-bar linkage in the rigid-deployable patterns forms a pair of straight lines in both contracted states, and hence there is no geometrical frustration throughout the deployment (see also SI Video~S2). The trajectory of the tiles throughout the deployment and the zero energy associated with deployment process confirms that the pattern morphs smoothly from a contracted configuration to a deployed configuration and subsequently to another contracted configuration. It is noteworthy that the reconfigurability constraints and the rigid-deployability constraints imply that all negative space rhombi are similar (i.e. congruent up to rescaling). Therefore, the change in the orientation of the tiles forms a checkerboard pattern with a magnitude that is spatially uniform and changes continuously from 0 to $\pi/2$ throughout the deployment and contraction. It is also possible to perform the constrained optimization directly on the two contracted configurations without caring about the intermediate states (see SI Section~S3 for more details). To realize this computation in a physical setting, we fabricated a model by laser-cutting acrylic plastic sheets and connecting them with tape joints (see also SI Video~S3 and Section~S6) and confirm that the physical model behaves just as predicted. As a striking example of this approach, we revisit the question of circling the square~\cite{choi2019programming} via a reconfigurable, rigidly-deployable kirigami pattern, with the results shown in Fig.~\ref{fig:F3}C (see also SI Video~S4--S5 and Section~S6), with the physical model made of laser-cut wooden tiles connected with tapes. Note that a major difference between the pattern in Fig.~\ref{fig:F3}C and the one in Fig.~\ref{fig:F2}A is that here the circle shape is achieved at the reconfigured contracted state, while in Fig.~\ref{fig:F2}A the circle shape is achieved at the deployed state. The results demonstrate the flexibility of our proposed reconfigurable kirigami design framework which yields geometric constructions that are material-independent (see SI Section~S5 for more results). 

While we have primarily focused on the quad kirigami patterns so far, other bases such as the kagome (triangle-based) kirigami patterns may also be used to construct reconfigurable kirigami patterns (see Section~S7 for details). It is also possible to extend our approach to produce three-dimensional kirigami-based reconfigurable tubular structures that morph from one contracted configuration into another contracted configuration (see Section~S8 for details). 

%So far, we have limited ourselves to planar kirigami patterns. We now show how to extend our approach to three-dimensional reconfigurable structures. To achieve these designs, we introduce cuts into a given target tubular shape into patches and isometrically unfold them onto the plane. We then apply our constrained optimization framework to produce a reconfigurable kirigami pattern for each planar shape, with the periodicity of the boundaries corresponding to the cuts enforced. Finally, the patterns are mapped back to the three-dimensional space to form a reconfigurable tubular structure. In Fig.~\ref{fig:F5}A, we show three-dimensional kirigami-based reconfigurable tubular structures that morph from one contracted configuration into another contracted configuration. Fig.~\ref{fig:F5}B shows another reconfigurable tubular structure achieved by this approach. When compared to the one in Fig.~\ref{fig:F5}A, this structure exhibits a smaller axial expansion but a more nonuniform radial change throughout the deployment and contraction process. More complex reconfigurable tubular structures can then be constructed using multiple copies of the patterns obtained by the optimization framework (Fig.~\ref{fig:F5}C). While we have primarily focused on the quad kirigami patterns so far, other bases such as the kagome (triangle-based) kirigami patterns may also be used to construct reconfigurable kirigami patterns (see Section~S7 for details). 

Our approach exploits the duality in kirigami patterns to create a class of reconfigurable, rigidly-deployable kirigami patterns with multiple closed and compact configurations. Just as flat-foldability and rigid-foldability of origami~\cite{kawasaki1989relation} opened the way for understanding and extending origami designs, perhaps their natural analogs in kirigami, viz. contractibility and rigid-deployability that we have uncovered here might pave the way for a range of art-inspired mathematics, science and engineering. %Furthermore, the highly nontrivial shape change as a consequence of the local rotation and reorganization of the kirigami tiles may remind one of cell rearrangements in biology, and so it will be interesting to explore the relation between them in our future work. 

%As the reconfigurability is considered at the level of every unit cell in this work, a natural next step would be to to extend our approach for achieving reconfigurability in a hierarchical manner. 

%{\bf Acknowledgments } This work was supported in part by the National Science Foundation under Grant no.~DMS-2002103 (G.P.T.C.), DMR-2011754 (L.M.), DMR-1922321 (L.M.) and EFRI-1830901 (L.M.), and the Harvard Quantitative Biology Initiative and the NSF-Simons Center for Mathematical and Statistical Analysis of Biology at Harvard, award number \#1764269 (G.P.T.C., L.M.). 

\bibliographystyle{Science_with_title}
\bibliography{reconfigurablebib}

\newpage 

\begin{FPfigure}
\centering
\includegraphics[width=\textwidth]{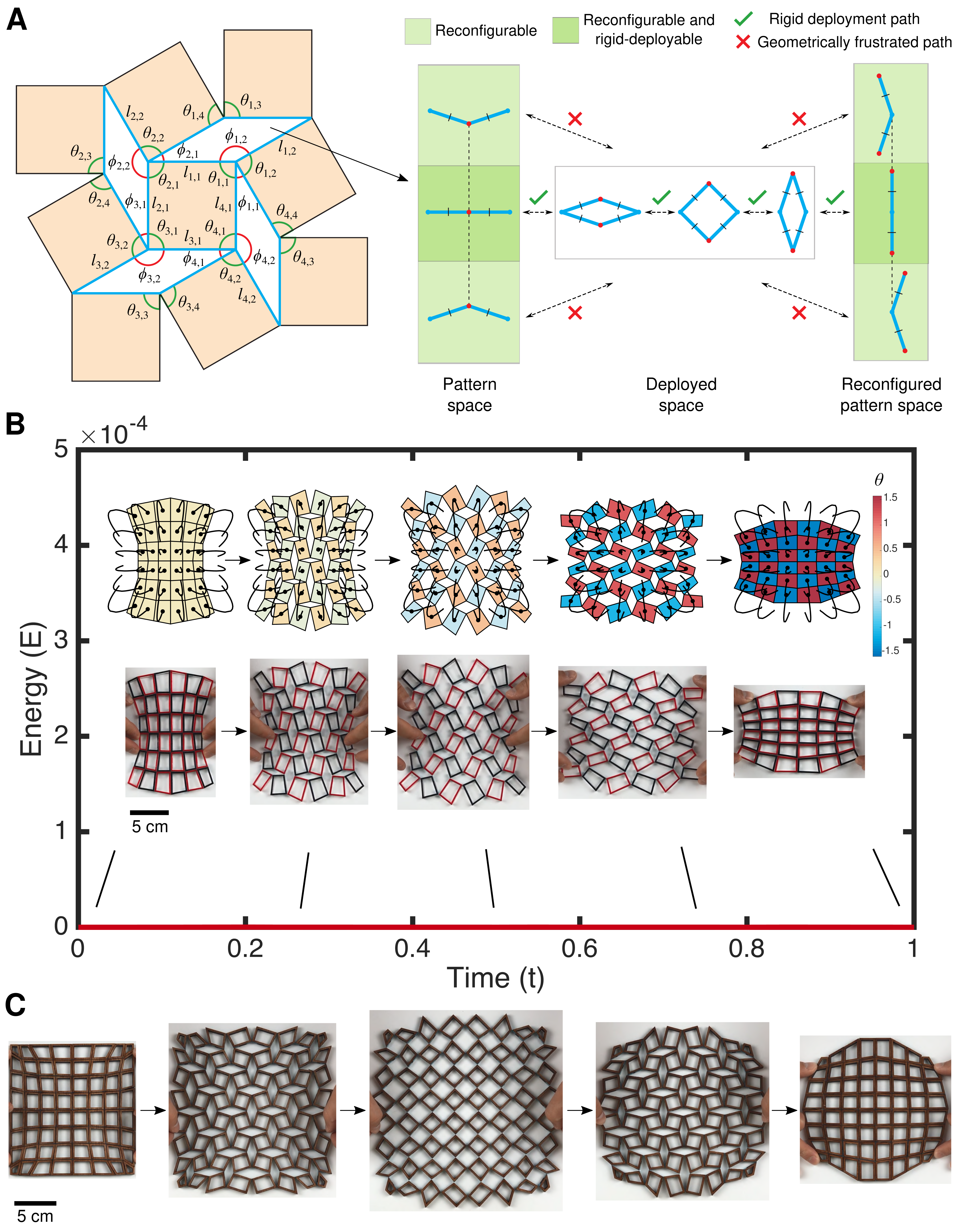}
\caption{\textbf{Reconfigurable, rigid-deployable kirigami.} (A)~(Left)~In kirigami patterns, each negative space (blue) formed by a generic deployed quad kirigami structure (not necessarily reconfigurable) is a four-bar linkage with two pairs of adjacent edges having the same length. (Right) If the reconfigurability constraints are enforced, all links in the four-bar linkage have the same length. Such a linkage has three rigid deployments (shown by the dotted lines), one non-trivial path in which all angles between links are activated (the horizontal dotted lines) and two degenerate paths connected by branch points at the ends of the first path (the vertical dotted lines). The linkage can deploy rigidly from the branch point into either deployment paths, but cannot rigidly transform directly between points on the deployment paths while remaining embedded in two dimensions. If Eq.~\eqref{eqt:straight} is satisfied, the four-bar linkage in the pattern space will be as shown in the dark green box, and hence there is a rigid deployment path (as indicated by the green ticks) for morphing it from the pattern space to the deployed space and then to the reconfigured pattern space. However, if Eq.~\eqref{eqt:straight} is not satisfied, the four-bar linkage in the pattern space will be as shown in the pale green boxes, and so there is no rigid deployment path for reconfiguring it (as indicated by the red crosses). (B) The deployment of a reconfigurable and rigid-deployable kirigami pattern, with snapshots of a numerical model and a physical acrylic plastic model at different states. Because of the rigid-deployability constraints, a flat energy landscape is observed. Each black curve in the numerical model shows the trajectory of a tile center, with the black dot indicating the current position of it. The color of each tile indicates the orientation change (denoted by $\theta$) of the tile with respect to the initial contracted state (see videos S2, S3 and SI for details). (C) Transforming the square to a circle - a physical wooden model of a reconfigurable and rigid-deployable kirigami pattern achieving a square-to-circle transformation through deployment and contraction (see videos S4, S5 and SI for details).}
\label{fig:F3}
\end{FPfigure}
%(Inset) Such a linkage has two one-dimensional rigid deployments connected by a single branch point, the configuration with all edges collinear and an angle of $\pi$ between overlapping edge pairs at the common hinge (red). The linkage can deploy rigidly from the branch point into either deployment paths, but cannot rigidly transform directly between points on the deployment paths while remaining embedded in two dimensions. 

%\begin{figure}[t!]
%\centering
%\includegraphics[width=\textwidth]{F5_compressed.pdf}
%\caption{\textbf{Reconfigurable tubular structures.} (A) Our framework can be extended for creating reconfigurable tubular structures which morph from one three-dimensional contracted configuration into another three-dimensional contracted configuration. Given a target 3D shape, we first cut and unfold it onto the plane. We then solve the 2D constrained optimization problem with an additional periodic boundary constraint to produce a reconfigurable kirigami pattern for the planar shape. We can then map the design back to 3D to form a reconfigurable tubular structure. (B) Another reconfigurable tubular structure with a more nonuniform radial change throughout the deployment and contraction process. (C) A more complex reconfigurable tubular structure produced by assembling four copies of the pattern in~(B).} 
%\label{fig:F5}
%\end{figure}

\end{document}